\RequirePackage{xcolor}
\documentclass[article,nojss]{jss}

\usepackage[utf8]{inputenc}
\usepackage[T1]{fontenc}

\usepackage{amsmath,amssymb,array}

\usepackage{algorithmicx}
\usepackage{algpseudocode}
\usepackage{algorithm}

\usepackage{pgfplotstable,filecontents}
\pgfplotsset{compat=1.9}

\usepackage{multirow}

\author{Anthony  Ebert\\Queensland University of Technology\\ACEMS \And 
        Paul Wu\\Queensland University of Technology\\ACEMS \AND
        Kerrie Mengersen\\Queensland University of Technology\\ACEMS \And
        Fabrizio Ruggeri\\CNR-IMATI\\Queensland University of Technology\\ACEMS}
\title{Computationally Efficient Simulation of Queues: The \proglang{R} Package \pkg{queuecomputer}}

\Shorttitle{queuecomputer: Computationally Efficient Simulation of Queues}

\Plainauthor{Anthony Ebert, Paul Wu, Kerrie Mengersen, Fabrizio Ruggeri} 
\Plaintitle{Computationally Efficient Simulation of Queues: The R Package queuecomputer} 

\Abstract{
Large networks of queueing systems model important real-world systems such as MapReduce clusters, web-servers, hospitals, call centers and airport passenger terminals. To model such systems accurately, we must infer queueing parameters from data. Unfortunately, for many queueing networks there is no clear way to proceed with parameter inference from data. Approximate Bayesian computation could offer a straightforward way to infer parameters for such networks if we could simulate data quickly enough. 

We present a computationally efficient method for simulating from a very general set of queueing networks with the \proglang{R} package \textbf{queuecomputer}. Remarkable speedups of more than 2 orders of magnitude are observed relative to the popular DES packages \pkg{simmer} and \pkg{simpy}. We replicate output from these packages to validate the package. 

The package is modular and integrates well with the popular \proglang{R} package \pkg{dplyr}. Complex queueing networks with tandem, parallel and fork/join topologies can easily be built with these two packages together. We show how to use this package with two examples: a call center and an airport terminal. 
}
\Plainkeywords{queues, queueing theory, discrete event simulation, 
operations research, approximate Bayesian computation, R} 

\Address{
  Anthony Ebert\\
  School of Mathematical Sciences\\
  Science and Engineering Faculty\\
  Queensland University of Technology\\
  Brisbane Queensland 4000, Australia\\
  E-mail: \email{ac.ebert@qut.edu.au}\\
  URL: \url{https://acems.org.au/our-people/anthony-ebert}
}

\begin{document}

\section{Introduction}

The queues we encounter in our everyday experience, where customers wait in line to be served by a server, are a useful analogy for many other processes. We say analogy because the word customers could represent: MapReduce jobs \citep{lin_joint_2013}; patients in a hospital \citep{takagi_queueing_2016}; items in a manufacturing system \citep{dallery_manufacturing_1992}; calls to a call center \citep{gans_telephone_2003}; shipping containers in a seaport \citep{kozan_comparison_1997} or even cognitive tasks \citep{cao_queueing_2013}. Similarly, server could represent: a compute cluster; medical staff; machinery or a customer service representative at a call center. Queueing systems can also be networked together to form queueing networks. We can use queueing networks to build models of processes such as provision of internet services \citep{sutton_bayesian_2011}, passenger facilitation at international airports \citep{wu_review_2013} and emergency evacuations \citep{van_woensel_modeling_2007}. Clearly queueing systems and queueing networks are useful for understanding important real-world systems. 

Performance measures for a given queueing system can often only be derived through simulation. Queues are usually simulated with discrete event simulation (DES) \citep[pg. 226]{insua2012bayesian}. In DES changes in state are discontinuous. The state is acted upon by a countable list of events at certain times which cause the discontinuities. If the occurrence of an event is independent of everything except simulation time it is determined; otherwise, it is contingent \citep{nance1981time}. 

Popular DES software packages are available in many programming languages including: the \proglang{R} package \pkg{simmer} \citep{Rpkg_simmer}, the \proglang{Python} \citep{van2011python} package \pkg{simpy} \citep{Ppkg_simpy} and the \proglang{Java}  \citep{gosling2000java} package \pkg{JMT} \citep{Jpkg_JMT}. DES packages are often so expressive that they can be considered languages in their own right, indeed the programming language \proglang{Simula} \citep{dahl1966simula} is a literal example of this. 

\pkg{queuecomputer} \citep{Rpkg_queuecomputer} implements an algorithm that can easily be applied to a wide range of queueing systems and networks of queueing systems. It is vastly more computationally efficient than existing approaches to DES. We term this new computationally efficient algorithm queue departure computation (QDC). Computational efficiency is important because if we can simulate from queues quickly, then we can embed a queue simulation within an approximate Bayesian computation (ABC) algorithm \citep{sunnaker_approximate_2013} and estimate queue parameters for very complicated queueing models in a straightforward manner. 

In Section \ref{sec:queueing} we review the literature on queueing theory and develop notation used throughout this paper. In Section \ref{sec:QDC} we present the QDC algorithm and compare it to DES. We demonstrate usage of the package in Section \ref{sec:usage}. Details of implementation and usage are discussed in Section \ref{sec:Implementation}. The package is validated in Section \ref{sec:Validation} by replicating results from DES packages \pkg{simpy} and \pkg{simmer}. We compare computed performance measures from the output of a \pkg{queuecomputer} simulation to theoretical results for $M/M/2$ queueing systems. We benchmark the package in Section \ref{sec:Benchmark} and compare computation time with \pkg{simpy} and \pkg{simmer}. Examples in Section \ref{sec:Examples} are used to demonstrate how the package can be used to simulate a call center and an international airport terminal. 

\section{Queueing theory} \label{sec:queueing}

Queueing theory is the study of queueing systems and originated from the work of Agner Krarup Erlang in 1909 to plan infrastructure requirements for the Danish telephone system \citep[pg 2]{thomopoulos2012fundamentals}. 

A queueing system is defined as follows. Each customer $i = 1,2,\cdots$ has an arrival time $a_i$ (or equivalently an inter-arrival time $\delta_i = a_i - a_{i-1}$, $a_0 = 0$) and an amount of time they require with a server, called the service time $s_i$. Typically a server can serve only one customer at a time. A server which is currently serving another customer is said to be unavailable, a server without a customer is available. If all servers are unavailable when a customer arrives then customers must wait in the queue until a server is available. Detailed introductions to queueing systems can be found in standard texts such as \citet{bhat2015introduction}. 

The characteristics of a queueing system are expressed with the notation of \citet{kendall1953stochastic}. This notation has since been extended to six characteristics:
\begin{itemize}
\item $f_{\mathbf{\delta}}$, inter-arrival distribution;
\item $f_{\mathbf{s}}$, service distribution; 
\item $K$, number of servers $\in \mathbb{N}$;
\item $C$, capacity of system $\in \mathbb{N}$; 
\item $n$, customer population $\in \mathbb{N}$; and
\item $R$, service discipline
\end{itemize}

Choices for inter-arrival and service distributions are denoted by ``M" for exponential and independently distributed, ``GI" for general and independently distributed and ``G" for general without the independence assumption. The capacity of the system $C$ refers to the maximum number of customers within the system at any one time\footnote{If the system is at full capacity and new customers arrive, new customers leave the system immediately without being served.}. Customers are within the system if they are being served or waiting in the queue. The customer population $n$ is the total number of customers including those outside of the system (yet to arrive or already departed). The service discipline $R$ defines how customers in the queue are allocated to available servers. The most common service discipline is first come first serve (FCFS). To specify a queueing system, these characteristics are placed in the order given above and separated by a forward slash ``/". 

The simplest queueing system is exponential in distribution for both the inter-arrival $\delta_i \overset{iid}{\sim} \mathrm{exp}(\lambda) ~ \forall i \in 1:n$ and service processes $s_i \overset{iid}{\sim} \mathrm{exp}(\mu) ~ \forall i \in 1:n$, where $\lambda$ and $\mu$ are exponential rate parameters. Additionally, $K$ is set to 1, $C$ and $n$ are infinite, and $R$ is FCFS. It is denoted by $M/M/1/ \infty / \infty / FCFS$, which is shortened to $M/M/1$. 

Parameter inference for this system was considered first by \citet{clarke1957maximum}, estimators were derived from the likelihood function. This likelihood is later used by \citet{muddapur1972bayesian} to derive the joint posterior distribution. Bayesian inference for queueing systems is summarised in detail by \citet{insua2012bayesian}. 

Managers and planners are less interested in parameter inference and more interested in performance measures such as: $N(t)$, the number of customers in system at time $t$; $\bar{B}$, the average number of busy servers; $\rho$, the resource utilization; and $\bar{\mathbf{w}}$, the average waiting time for customers. If $\lambda < K \mu$ the queueing system will eventually reach equilibrium and distributions of performance measures become independent of time. 

In the case of a $M/M/K$ system equilibrium distributions for performance measures are derived analytically, they are found in standard queueing theory textbooks \citep{lipsky2008queueing, thomopoulos2012fundamentals}. For instance, the limit probability of $N$ customers in the system $\Prob(N)$  is

\begin{align}
\Prob(0) &= \left[\frac{ (K \rho)^K}{K! (1 - \rho)} + 1 + \sum_{i=1}^{K-1} \frac{(K \rho)^i}{i!}   \right]^{-1} \nonumber \\
\Prob(N) &= \begin{cases} 
\Prob(0) \frac{(K \rho)^n }{N!} & \quad  N \leq K\\
\Prob(0) \frac{(K \rho)^n }{K! K^{N-K}} & \text{otherwise} 
\end{cases} \label{eq:pn}
\end{align}

where $\rho$, the resource utilization, is defined as $\frac{\lambda}{K \mu}$. For an $M/M/K$ system this is equal to the expected number of busy servers divided by the total number of servers $\frac{\E(B)}{K}$ \citep[pg. 451]{cassandras2009introduction}. The expected number of customers in the system is \citep{bhat2015introduction}  

\begin{align}
\E(N) = K\rho + \frac{\rho (K\rho)^K P(0)}{K!(1-\rho)^2} \label{eq:en},
\end{align}

and the expected waiting time is

\begin{align}
\E(\mathbf{w}) = \frac{(K \rho)^K P(0)}{K! K \mu(1-\rho)^2} \label{eq:ew}.
\end{align}

If the parameters of $f_\delta$ and $f_{\mathbf{s}}$ are uncertain, then we must turn to predictive distributions for estimates of performance measures, which are computed analytically for M/M/K queues (Equations \ref{eq:en} and \ref{eq:ew}). Predictive distributions of performance measures using Bayesian posterior distributions are derived by \citet{armero1994bayesian, armero1999dealing}.  

\citet{jackson_networks_1957} was one of the first to consider networks of queueing systems. In a Jackson network, there is a set of $J$ queueing systems. After a customer is served by queueing system $j$, they arrive at another queueing system with fixed probability $p_{j,k}$. Customers leave the system with probability $1 - \sum_{k=1}^{J} p_{j,k}$. Other examples of queueing networks include the tandem \citep{glynn_departures_1991}, parallel \citep{hunt_fast_1995} and the fork/join \citep{kim_analysis_1989} topologies. 

In a tandem queueing network, customers traverse an ordered series of queues before departing the system. Real examples of such systems include airport terminals, internet services and manufacturing systems. In a parallel network, customers are partitioned into different $\mathbf{(a,s)}$ to be seen by separate queueing systems. In a fork/join network a task (another term for customer) is forked into a number of subtasks which are to be completed by distinct parallel servers. The difference from the parallel network is that the task can only depart the system once all subtasks have arrived at the join point. 

Most models of queueing systems assume time-invariant inter-arrival and service processes. In practice, many real-world queues have inter-arrival processes which are strongly time-dependent, such as: call centers \citep{weinberg2007bayesian, brown2005statistical}, airport runways \citep{koopman1972air} and hospitals \citep{brahimi_queueing_1991}. In the case of the $M/M/1$ queue, we can adapt the notation to $M(t)/M(t)/1$ to represent exponential  processes where parameters $\lambda(t)$ and $\mu(t)$ change with time. Such queueing systems are referred to as dynamic queueing systems.  

In general, analytic solutions do not exist for dynamic queueing systems \citep{malone1995dynamic, worthington2009reflections}. \citet{green1991some} showed that using stationary queueing systems to model dynamic queueing systems leads to serious error even if deviation from stationarity is slight. The problem is compounded once we consider queueing networks. Understanding long-term and transient behaviour of such queues can only be achieved with approximation methods or simulation. We now detail the QDC algorithm, a computationally efficient method for simulating queueing systems. 

\section{Queue departure computation} \label{sec:QDC}

\subsection{Fixed number of servers}

QDC can be considered as a multiserver extension to an algorithm presented by \citet{lindley_theory_1952}. For a single server queueing system, the departure time of the $i$th customer is: $d_i = \max{(a_i, d_{i-1})} + s_i$, since the customer either waits for a server or the server waits for a customer. The algorithm (not the paper) was, surprisingly, not extended to multiserver systems until \citet{krivulin_recursive_1994}. However with each new customer $i$ the algorithm must search a growing $i+1$ length vector. This algorithm, therefore, scales poorly, with computational complexity $O(n^2)$, where $n$ is number of customers. \citet{kin_generalized_2010} adapted the original algorithm of \citet{kiefer_theory_1955} to an $O(nK)$ algorithm for multiserver tandem queues with blocking, that is $G/G/K/C$ queueing systems where $C$ is the maximum capacity number of customers in the queueing systems. 

QDC can also be viewed as a computationally efficient solution to the set of equations presented in \citet[pg. 259]{sutton_bayesian_2011} for FCFS queueing systems. There is a single queue served by a fixed number of $K$ servers. The $i$th customer observes a set of times $b_i = \{ b_{ik} | k \in 1:K \}$ which represents the times when each server will next be available. The customer $i$ selects the earliest available server $p_i = \text{argmin}(b_i)$ from $b_i$. The departure time for the $i$th customer is, therefore, $d_i = \max({a_i, b_{p_i}}) + s_i$, since the server must wait for the customer or the customer must wait for the server. The QDC algorithm for a fixed number of servers (Algorithm \ref{alg:fixed_servers}) pre-sorts the arrival times. Rather than assigning a $b_i$ for each customer $i$ to form the matrix $\mathbf{b} \in \mathbf{M}^{n \times K}$, QDC considers $\mathbf{b}$ as a continually updated $K$ length vector representing the state of the system. 

\begin{algorithm}[ht!]
\caption{QDC for fixed $K$}
\label{alg:fixed_servers}
\begin{algorithmic}[1]
\Function{QDC\_numeric}{$\mathbf{a} \in \mathbb{R}_{+}^{n}, \mathbf{s} \in \mathbb{R}_{+}^{n}, K \in \mathbb{N}$}
\State Sort $(\mathbf{a}, \mathbf{s})$ in terms of $\mathbf{a}$ (ascending)
\State Create vector $\mathbf{p} \in \mathbb{N}^{n}$.
\State Create vector $\mathbf{b} \in \mathbb{R}_{+}^{K}$. 
\State Create vector $\mathbf{d} \in \mathbb{R}_{+}^{n}$.
\State $b_k \leftarrow 0 \quad \quad \forall k \in 1:K $
\For{$i \in 1:n$}
  \State $p_i \leftarrow \text{arg min}(\mathbf{b}) $ \label{alg_line:search}
  \State $b_{p_i} \leftarrow \max(a_i, b_{p_i}) + s_i $
  \State $d_i \leftarrow b_{p_i}$ 
\EndFor 
\State Put ($\mathbf{a}, \mathbf{d}, \mathbf{p}$) back to original (input) ordering of $\mathbf{a}$ 
\State \Return ($\mathbf{d}, \mathbf{p}$)
\EndFunction
\end{algorithmic}
\end{algorithm}

This algorithm is simple and computationally efficient. At each iteration of the loop, we need only search $\mathbf{b}$, a $K$ length vector for the minimum element in code line \ref{alg_line:search}. In the language of DES, we consider $\mathbf{b}$ as the system state and $\mathbf{a}$ as the event list, which are all determined events. This differs from conventional DES approaches to modelling queueing systems where the queue length is the system state, and both $\mathbf{a}$ and $\mathbf{d}$ constitute the event list, where the events of $\mathbf{a}$ are determined and the events of $\mathbf{d}$ are continually updated and therefore contingent.  

Algorithm \ref{alg:fixed_servers} can simulate any queue of the form $G(t)/G(t)/K/\infty/n/FCFS$ where $K$ and $n$ can be made arbitrarily large. Furthermore, the inter-arrival and service distributions can be of completely general form and even have a dependency structure between them. Since the arrival and service times are supplied by the user rather than sampled in-situ, the algorithm ``decouples" statistical sampling from queue computation. This frees the user to simulate queues of arbitrarily complex $f_{\delta, \mathbf{s}}$, where $K$ is fixed. 

\subsection{Changing number of servers}

\subsubsection{Conditional case}

Suppose that the number of servers that customers can use changes throughout the day. This reflects realistic situations where more servers are rostered on for busier times of the day. We say that for a certain time $t$, the customers have a choice of $K(t)$ open servers from $K$. This means that there are $K(t)$ servers rostered-on for time $t$. We define the term closed as the opposite of open. 

We represent the number of open servers throughout the day as a step function. Time is on the positive real number line and is partitioned by $L$ knot locations $\mathbf{x} = (x_1, \cdots, x_L ) \in \mathbb{R}^L_{+}$ into $L+1$ epochs $(0, x_1], (x_1, x_2], \cdots, (x_L, \infty)$. The number of open servers in each epoch is represented by a $L+1$ length vector $\mathbf{y} = (y_1, \cdots, y_{L+1} ) \in \mathbb{N}^{L+1}_0$. If we assume that none of the service times $\mathbf{s}$ span the length of more than one epoch $(x_{l}, x_{l+1}]$, formally

\begin{align}
\forall i \quad \left[ s_i < \min({ x_{l+1} - x_{l} | l \in 1:L }) \right] \label{eq:condition},
\end{align}

then we need to consider a change in state over at most 1 knot location. This step function is determined input by the user. Like the arrival and service times $\mathbf{(a,s)}$ it is changeable by the user before the simulation but not during the simulation. 

We close server $k$ by writing an $\infty$ symbol to $b_k$ ensuring that no customer can use that server. If the server needs to be open again at time $t$, we write $t$ to $b_k$ allowing customers to use that server. Since $\mathbf{x}$ now corresponds to changes in $\mathbf{b}$, it is part of the event list along with $\mathbf{a}$. The entire event list is still determined and need not be updated mid-simulation. 

\begin{algorithm}[!ht]
\caption{QDC for $K(t)$ (conditional)}
\label{alg:changing_servers}
\footnotesize
\begin{algorithmic}[1]
\Function{QDC\_server.stepfun}{$\mathbf{a} \in \mathbb{R}_{\text{+}}^{n}, \mathbf{s} \in \mathbb{R}_{+}^{n}, \mathbf{x} \in \mathbb{R}^{L}_{+}, \mathbf{y} \in \mathbb{N}^{L+1}_0 $}
\State Sort $(\mathbf{a}, \mathbf{s})$ in terms of $\mathbf{a}$ (ascending)
\State $x_{L+1} \leftarrow \infty$
\State $y_{L+2} \leftarrow 1$
\State $K \leftarrow \max(\mathbf{y})$
\State Create vector $\mathbf{b} \in \mathbb{R}_{\text{+}}^{K}$. 
\State $b_k \leftarrow \infty \quad \quad \forall k \in 1:K $
\State $b_k \leftarrow 0 \quad \quad \forall k \in 1:y_0 $
\State Create vector $\mathbf{p} \in \mathbb{N}^{n}$.
\State Create vector $\mathbf{d} \in \mathbb{R}_{+}^{n}$.
\State $l \leftarrow 1$
\State $p_1 \leftarrow 1$ 
\For{$i \in 1:n$} \\ \\

\hspace{10mm} \textit{// Adjustments to $\mathbf{b}$ with change in epoch}
  \If{$\forall k \in 1:K \quad \left[ b_k \geq x_{l+1} \right]$ OR $a_i \geq x_{l+1}$}
    \If{$ y_{l+1} - y_l > 0 $}
      \For{$k \in (y_l + 1 : y_{l+1}) $}
      	\State $b_k \leftarrow x_{l+1}$ 
      \EndFor
    \EndIf
    \If{$ y_{l+1} - y_l < 0 $}
      \For{$k \in (y_{l+1} + 1 : y_{l}) $}
        \State $b_k \leftarrow \infty$ 
      \EndFor
    \EndIf
    \State $l \leftarrow l + 1$
  \EndIf \\ 
 \hspace{10mm} \textit{// End of adjustments to $\mathbf{b}$ with change in epoch} \\
  \State $p_i \leftarrow \text{arg min}(\mathbf{b}) $
  \State $b_{p_i} \leftarrow \max(a_i, b_{p_i}) + s_i $
  \State $d_i \leftarrow b_{p_i}$ \\ \\
\hspace{10mm} \textit{// Extra loop if current size is zero so that customer $i$ can be processed in next epoch}
  \If{$y_l = 0$}
  	\State $i \leftarrow i - 1$
  \EndIf \\
\EndFor 
\State Put ($\mathbf{a}, \mathbf{d}, \mathbf{p}$) back to original (input) ordering of $\mathbf{a}$ 
\State \Return ($\mathbf{d}, \mathbf{p}$)
\EndFunction
\end{algorithmic}
\end{algorithm}

This algorithm can simulate queues of form $G(t)/G(t)/K(t)/\infty/n/FCFS$, where $K(t)$ refers to the number of open servers changing with time. As mentioned previously this algorithm is subject to Condition \ref{eq:condition}. This condition is not overly restrictive if we consider realistic systems with few changes in $K$. The recorded server allocations $\mathbf{p} = (p_1, \cdots, p_n)$ may not reflect the real system since Algorithm \ref{alg:changing_servers} does not allow the user to specify exactly which servers are open in each epoch, only how many are open and closed. If this output is needed or in cases where Condition \ref{eq:condition} does not hold, we must use the less computationally efficient but more general unconditional algorithm below.  

\subsubsection{Unconditional case}

If Condition \ref{eq:condition} does not hold or if, otherwise, we wish to control exactly which servers are open at what time then we must use a less computationally efficient algorithm (Algorithm \ref{alg:changing_servers_unconditional}). Each server $k$ has its own partition of $L_k$ knot locations $\mathbf{x_k} = (x_{k,1}, \cdots, x_{k,L_k} ) \in \mathbb{R}^{L_k}_{+}$ and each $\mathbf{y_k} = (y_{k,1}, \cdots, y_{k,L_k+1} )$ is an alternating sequence of 0 and 1s of length $L_k$ indicating whether the server is open or closed respectively for the associated epoch. The vector $\mathbf{c}$ is used slightly differently to how it is used in \citet{sutton_bayesian_2011}. We use it to represent the time at which each server is next available for the current customer $i$, given the current system state $\mathbf{b}$. It is the output of the \code{next\_fun} function. 

\begin{algorithm}[ht!]
\caption{Next function}
\label{alg:next_fun}
\begin{algorithmic}
\Function{next\_fun}{$t, \mathbf{x} \in \mathbb{R}^{L}_{+}, \mathbf{y}$}
\State Find $l$ such that $x_{l} < t \leq x_{l+1}$.
\If{$y_{l+1} = 0$}
\State \Return $x_{l+1}$
\Else
\State \Return $t$
\EndIf
\EndFunction
\end{algorithmic}
\end{algorithm}

\begin{algorithm}[ht!]
\caption{QDC for $K(t)$ (unconditional)}
\label{alg:changing_servers_unconditional} 
\begin{algorithmic}[1]
\Function{QDC\_server.list}{$\mathbf{a} \in \mathbb{R}_{+}^{n}, \mathbf{s} \in \mathbb{R}_{+}^{n}, \mathbf{\underline{x}} = (\mathbf{x_1}, \cdots, \mathbf{x_K} ), \mathbf{\underline{y}} = (\mathbf{y_1}, \cdots, \mathbf{y_K} ) $}
\State Sort $(\mathbf{a}, \mathbf{s})$ in terms of $\mathbf{a}$ (ascending)
\State $\forall k \in 1:K \quad x_{k,L_k+1} \leftarrow \infty$
\State $\forall k \in 1:K \quad y_{k,L_k+2} \leftarrow 1$
\State $K \leftarrow $ length($\mathbf{\underline{x}}$)
\State Create vector $\mathbf{c} \in \mathbb{R}_{\text{+}}^{K}$
\State Create vector $\mathbf{b} \in \mathbb{R}_{\text{+}}^{K}$
\State $b_k \leftarrow 0 \quad \forall k \in 1:K$
\State Create vector $\mathbf{p} \in \mathbb{N}^{n}$.
\State Create vector $\mathbf{d} \in \mathbb{R}_{+}^{n}$.
\For{$i \in 1:n$}
	\For{$k \in 1:K$}
        \State $c_k \leftarrow$ \Call{next\_fun}{$\max(b_k, a_i), \mathbf{x_k}, \mathbf{y_k}$}
    \EndFor
\State $p_i \leftarrow \text{arg min}(\mathbf{b}) $
\State $b_{p_i} \leftarrow c_{p_i} + s_i $
\State $d_i \leftarrow b_{p_i}$ 
\EndFor 
\State Put ($\mathbf{a}, \mathbf{d}, \mathbf{p}$) back to original (input) ordering of $\mathbf{a}$ 
\State \Return ($\mathbf{d}, \mathbf{p}$)
\EndFunction
\end{algorithmic}
\end{algorithm}

This algorithm can simulate queueing systems of form $G(t)/G(t)/K(t)/\infty/n/FCFS$, where $K(t)$ refers to the number of open servers changing with time. In addition, we can specify which particular servers are available when, not just how many and we are not bound by Condition \ref{eq:condition}. Once again we note that $\mathbf{b}$ can be considered as the system state and the event list is formed by $\mathbf{a}$ and the elements of $\mathbf{\underline{x}}$. This function can be called with the \code{queue\_step} function in \pkg{queuecomputer} by supplying a \code{server.list} object to the \code{servers} argument. For the rest of this paper we focus on Algorithms \ref{alg:fixed_servers} and \ref{alg:changing_servers} for their relative conceptual simplicity and computational efficiency.

\subsection{Discussion}

With the algorithms so far presented, we can simulate from a very general set of queueing systems $G(t)/G(t)/K(t)/\infty/n/FCFS$ in a computationally efficient manner. In contrast to the algorithm of \cite{kin_generalized_2010}, the state vector $\mathbf{b}$ is written over in each iteration. The memory usage for QDC, therefore, scales with $O(n)$ rather than $O(nK)$. 

Tandem queueing networks can be simulated by using the output of one queueing system as the input to the next queueing system. We demonstrate this idea with the Airport Simulation examples in Section \ref{ssec:largerairport}. Fork/join queueing networks are addressed in the next section where we explain the implementation details of \pkg{queuecomputer} with regards to the QDC algorithm.  

\newpage

\section{Usage} \label{sec:usage}

The purpose of the package \pkg{queuecomputer} is to compute, deterministically, the output of a queueing system given the arrival and service times for all customers. The most important function is \code{queue\_step}. The first argument to  \code{queue\_step} is a vector of arrival times, the second argument is a vector of service times and the third argument specifies the servers available. 

\begin{CodeChunk}
\begin{Sinput}
R> library("queuecomputer")
R> arrivals <- cumsum(rexp(100))
R> head(arrivals)
\end{Sinput}
\begin{Soutput}
[1] 0.693512 1.693399 2.425550 3.952405 3.961906 4.405492
\end{Soutput}
\begin{Sinput}
R> service <- rexp(100)
R> departures <- queue_step(arrivals, service = service, servers = 2)
R> departures
\end{Sinput}
\begin{Soutput}
# A tibble: 100 × 6
   arrivals     service departures      waiting system_time server
      <dbl>       <dbl>      <dbl>        <dbl>       <dbl>  <dbl>
1  0.693512 0.830158956   1.523671 0.000000e+00 0.830158956      1
2  1.693399 0.817648174   2.511047 1.110223e-16 0.817648174      2
3  2.425550 0.002675641   2.428226 2.138047e-16 0.002675641      1
4  3.952405 0.667180991   4.619586 4.440892e-16 0.667180991      1
5  3.961906 0.551920432   4.513827 4.440892e-16 0.551920432      2
6  4.405492 1.069236762   5.583063 1.083341e-01 1.177570886      2
7  4.594253 1.110448926   5.730035 2.533279e-02 1.135781711      1
8  4.993053 0.766944956   6.350008 5.900099e-01 1.356954853      2
9  6.047412 0.805061421   6.852474 1.110223e-16 0.805061421      1
10 6.856338 1.317802131   8.174140 0.000000e+00 1.317802131      2
# ... with 90 more rows
\end{Soutput}
\end{CodeChunk}

The output of a \code{queue\_step} function is a \code{queue\_list} object. We built a summary method for objects of class \code{queue\_step}, which we now demonstrate. 

\begin{CodeChunk}
\begin{Sinput}
R> summary(departures)
\end{Sinput}
\begin{Soutput}
Total customers:
 100
Missed customers:
 0
Mean waiting time:
 0.246
Mean response time:
 1.11
Utilization factor:
 0.53
Mean queue length:
 0.301
Mean number of customers in system:
 1.36
\end{Soutput}
\end{CodeChunk}

If the last element of $\mathbf{y}$ is zero, it is possible that some customers will never be served, this is the ``Missed customers'' output. The performance measures that follow are the mean waiting time $\bar{w}$, the mean response time $\bar{r} = d - a$, the observed utilization factor $\bar{B}/K$, the mean queue length and the mean number of customers in the system respectively. The utilization factor $\bar{B}/K$ takes into account the changing number of open servers $K(t)$ where Algorithm \ref{alg:changing_servers} is used. We now explain the implementation details of package.

\section{Implementation} \label{sec:Implementation}

The \code{for} loops within Algorithms \ref{alg:fixed_servers} and \ref{alg:changing_servers} are written in \proglang{C++} with the \pkg{Armadillo} library \citep{sanderson2016armadillo}. The \proglang{C++} \textbf{for} loops are called using the \proglang{R} packages \pkg{Rcpp} \citep{eddelbuettel2011rcpp} and \pkg{RcppArmadillo} \citep{eddelbuettel2014rcpparmadillo}. We use \proglang{R} to provide wrapper functions for the \proglang{C++} code. 

The \code{queue\_step} calls the more primitive \code{queue} function which is a wrapper for S3 methods which implement Algorithms \ref{alg:fixed_servers}, \ref{alg:changing_servers} or \ref{alg:changing_servers_unconditional} depending on the class of the object supplied to the \code{server} argument of \code{queue\_step}. If \code{class(server)} is numeric, then \code{queue} runs Algorithm \ref{alg:fixed_servers}, if it is a \code{server.stepfun} then \code{queue} runs Algorithm \ref{alg:changing_servers}, if it is a \code{server.list} then \code{queue} runs Algorithm \ref{alg:changing_servers_unconditional}. The \code{queue} function computes departure times $\mathbf{d}$ and server allocations $\mathbf{p}$ and the \code{queue\_step} function adds additional output such as waiting times and queue lengths which are used in summary and plot methods.  

To simulate fork/join networks, the \pkg{queuecomputer} function \code{wait\_step} provides a simple wrapper to the \pkg{base} function \code{pmax.int}, this function computes the maximum of each row for a set of two equal length numeric vectors. The vectors represent the departure times for each subjob and the departure time for the entire job is the maximum of each subjob. 

In \pkg{simmer} and \pkg{simpy} users supply generator functions for simulating $\mathbf{\delta}$ and service times $\mathbf{s}$, the user enters the set of input parameters $\theta_I$ for these generator functions and starts the simulation. The inter-arrival time is resampled after each arrival and the service time is sampled when the server begins with a new customer. This makes it difficult to model queues where distributions for inter-arrival times do not make sense: like the immigration counter for an airport, where multiple flights generate customers; or when arrival times and service times are not independent. In \pkg{queuecomputer} sampling is ``decoupled'' from computation, the user samples $\mathbf{a}$ and service times $\mathbf{s}$ using any method. The outputs $\mathbf{d}$ and $\mathbf{p}$ are then computed deterministically. 

We now demonstrate the validity of \pkg{queuecomputer}'s output by replicating results from the DES packages \pkg{simmer} and \pkg{simpy}. We then replicate equilibrium analytic results of performance measures for the $M/M/2$ queue. 

\section{Validation} \label{sec:Validation}

\subsection[Comparison with simmer and simpy]{Comparison with \pkg{simmer} and \pkg{simpy}}

To demonstrate the validity of the algorithm we consider a $M/M/2/\infty/1000/FCFS$ queue. If QDC is valid for any $M/M/K$ queueing system, then it is valid for any $G(t)/G(t)/K$ queueing system. This is because any non-zero $\mathbf{(a,s)}$ could conceivably come from two exponential distributions, even if the probability of the particular realization is vanishingly small. We replicate exact departure times computed with the \pkg{simmer} and \pkg{simpy} packages using \pkg{queuecomputer}. First, we generate $\mathbf{a}$ and $\mathbf{s}$ to be used as input to all three packages. 

\begin{Code}
R> set.seed(1)
R> n_customers <- 10^4
R> lambda_a <- 1/1
R> lambda_s <- 1/0.9
R> interarrivals <- rexp(n_customers, lambda_a)
R> arrivals <- cumsum(interarrivals)
R> service <- rexp(n_customers, lambda_s)
\end{Code}

We now input these objects into the three scripts using \pkg{queuecomputer}, \pkg{simmer}, or \pkg{simpy}. First, we run the \pkg{queuecomputer} script. The \code{queuecomputer_output} object is sorted in ascending order so that the departure times can be compared to the DES packages.

\begin{CodeChunk}
\begin{Sinput}
R> queuecomputer_output <- queue_step(arrivals = arrivals, 
+    service = service, servers = 2)
R> head(sort(depart(queuecomputer_output)))
\end{Sinput}
\begin{Soutput}
[1] 1.340151 2.288112 2.639976 2.796572 3.249794 5.714967
\end{Soutput}
\end{CodeChunk}

The DES packages \pkg{simmer} and \pkg{simpy} are not built to allow users to input $(\mathbf{a,s})$ directly. Rather, the user supplies parameters for $f_{\delta}$ and $f_{\mathbf{s}}$ so that inter-arrival and service times can be sampled at each step when needed. To allow \pkg{simmer} and \pkg{simpy} to accept presampled input $(\mathbf{a,s})$ we use generator functions instead of \code{rexp(rate)} or \code{random.expovariate(rate)} calls in \proglang{R} and \proglang{Python} respectively, details of this work can be found in the supplementary material. We create an interface to \pkg{simmer} so that it can be called in the same way as \pkg{queuecomputer}. 

\begin{CodeChunk}
\begin{Sinput}
R> simmer_output <- simmer_step(arrivals = arrivals, 
+    service = service, servers = 2)
R> head(simmer_output)
\end{Sinput}
\begin{Soutput}
[1] 1.340151 2.288112 2.639976 2.796572 3.249794 5.714967
\end{Soutput}
\end{CodeChunk}

The same departure times are observed. Similarly in \proglang{Python} we create an interface to \pkg{simpy} so that it can be called in a similar way to \pkg{queuecomputer}. 

\begin{CodeChunk}
\begin{Sinput}
python> simpy_step(interarrivals, service)[0:6]
\end{Sinput}
\begin{Soutput}
array([ 1.34015149,  2.28811237,  2.63997568,  2.79657232,  3.24979406, 
	5.7149671 ])
\end{Soutput}
\end{CodeChunk}

A check of all three sorted vectors of $\mathbf{d}$ from each package revealed that all were equal to within 5 significant figures for every $d_i, i = 1:1000$.  

\subsection{Replicate theoretical results for M/M/3}

We use a $M/M/3/\infty/5 \times 10^6/FCFS$ simulation in \pkg{queuecomputer} to replicate theoretical equilibrium results for key performance indicators for a $M/M/2/\infty/\infty/FCFS$ queueing system. We set $\lambda$ to 1 and set $\mu$ to 2.

\subsubsection{Theoretical results}

We first note that the traffic intensity is $\rho$ of $2/3 = 0.\dot{6}$, which should correspond to the average number of busy servers. The probability of $N$ customers in the system is given by Equation~\ref{eq:pn}. We perform this computation up to $N = 20$ and display the results in Figure~\ref{fig:theoretical}. The expected waiting time is $\E(\mathbf{w})$ is $0.\dot{4}$ and the expected number of customers in the system $\E(N)$ is $2.\dot{8}$.

\subsubsection{Simulation results}

The inputs $\mathbf{a}$ and $\mathbf{s}$ must first be generated. 

\begin{CodeChunk}
\begin{Sinput}
R> set.seed(1) 
R> n_customers <- 5e6
R> lambda <- 2
R> mu <- 1
R> interarrivals <- rexp(n_customers, lambda)
R> arrivals <- cumsum(interarrivals)
R> service <- rexp(n_customers, mu)
R> K = 3
\end{Sinput}
\end{CodeChunk}

We now use the \code{queue\_step} function and the summary method for \code{queue\_list} objects \code{summary.queue\_list} to return observed key performance measures. 

\begin{CodeChunk}
\begin{Sinput}
R> MM3 <- queue_step(arrivals = arrivals, service = service, servers = K)
R> summary(MM3)
\end{Sinput}
\begin{Soutput}
Total customers:
 5000000
Missed customers:
 0
Mean waiting time:
 0.445
Mean response time:
 1.44
Utilization factor:
 0.666140156160826
Mean queue length:
 0.889
Mean number of customers in system:
 2.89
\end{Soutput}
\end{CodeChunk}

\begin{figure}[!htb]
  \centering
  \includegraphics[width = 0.5\textwidth]{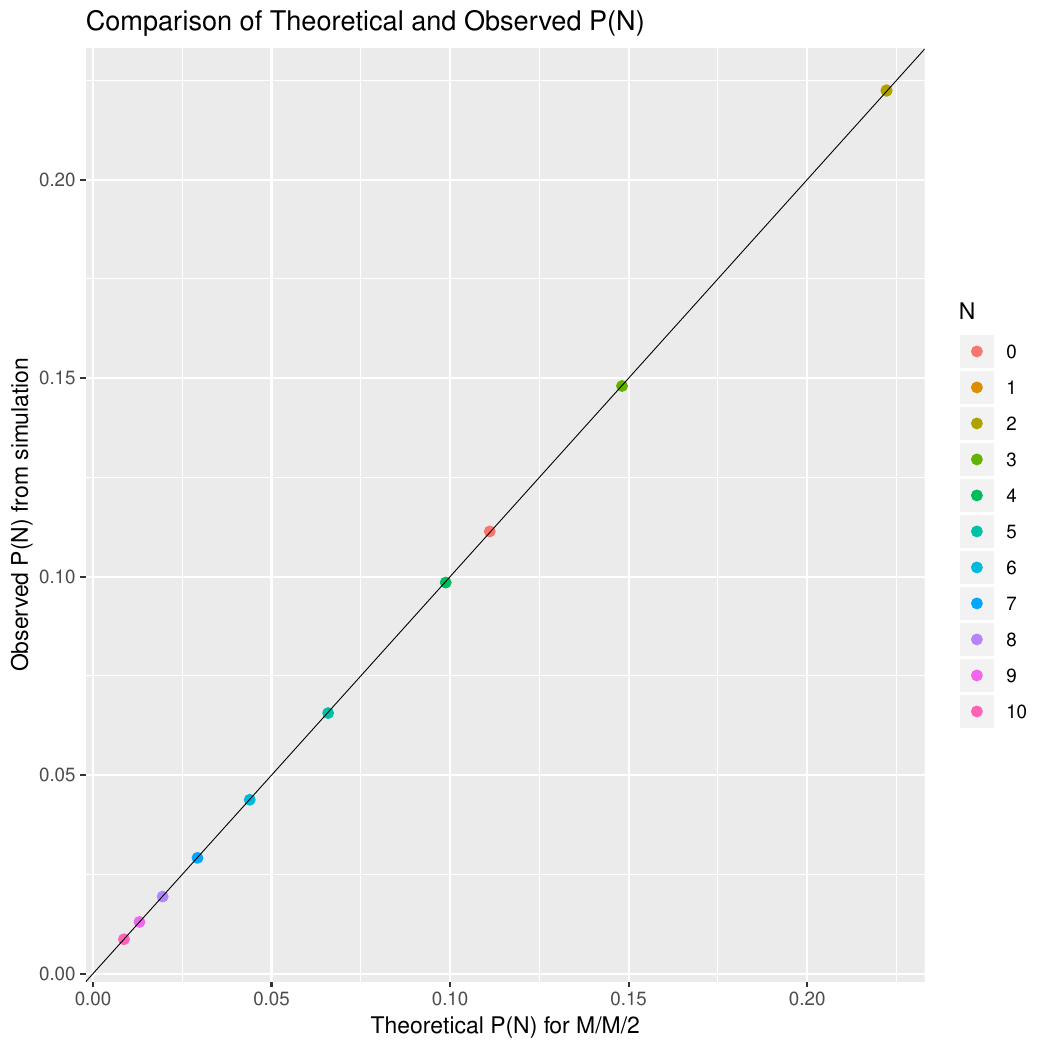}
  \caption{Comparison of theoretical equilibrium $\Prob(N)$ and observed proportions from simulation. Observation $N = 1$ is obscured by $N = 2$.}
  \label{fig:theoretical}
\end{figure}

We see that the observed time average number of busy servers is 0.6661402 which is close to $0.\dot{6}$ the value for $\rho$. We can see that the observed mean waiting time is close to the expected mean waiting time. The expected number of customers in the system, from the distribution $\Prob(N)$ is close to the observed number of customers in the system. The entire distribution of $\Prob(N)$ is replicated in Figure~\ref{fig:theoretical}. 

\section{Benchmark} \label{sec:Benchmark}

\subsection{Method}

The compare the computational efficiency of each package we compute the departure times from a $M/M/2/\infty/n/FCFS$ queueing system, with $\lambda = 1$ and $\mu = 1.\dot{1}$. To understand how $n$ affects computation time we repeat the experiment 100 times for n = $ 10^2, 10^3, 10^5$ and $10^6 $. We also repeat the experiment at $n = 10^7$ for \pkg{queuecomputer}. We compare the median time taken for each combination of package and $n$. 

The simulation was conducted on a system with Intel (R) Core(TM) i7-6700 CPU @ 3.40GHz running Debian GNU/Linux. The version of \proglang{R} is 3.5.1 ``Feather spray" with \pkg{simmer} version 4.0.1 and \pkg{queuecomputer} version 0.8.3. The version of \proglang{Python} is 3.5.3 with \pkg{simpy} module version 3.0.11.

To assess the computation time for \pkg{queuecomputer} and \pkg{simmer} we use the \code{microbenchmark} function from the \pkg{microbenchmark} package \citep{Rpkg_microbenchmark} with \code{time = 100} and compute the median. Full details can be found in the supplementary material. 

\subsection{Results and discussion}

\begin{figure}[!htb]
  \centering
  \includegraphics[width = 0.7\textwidth]{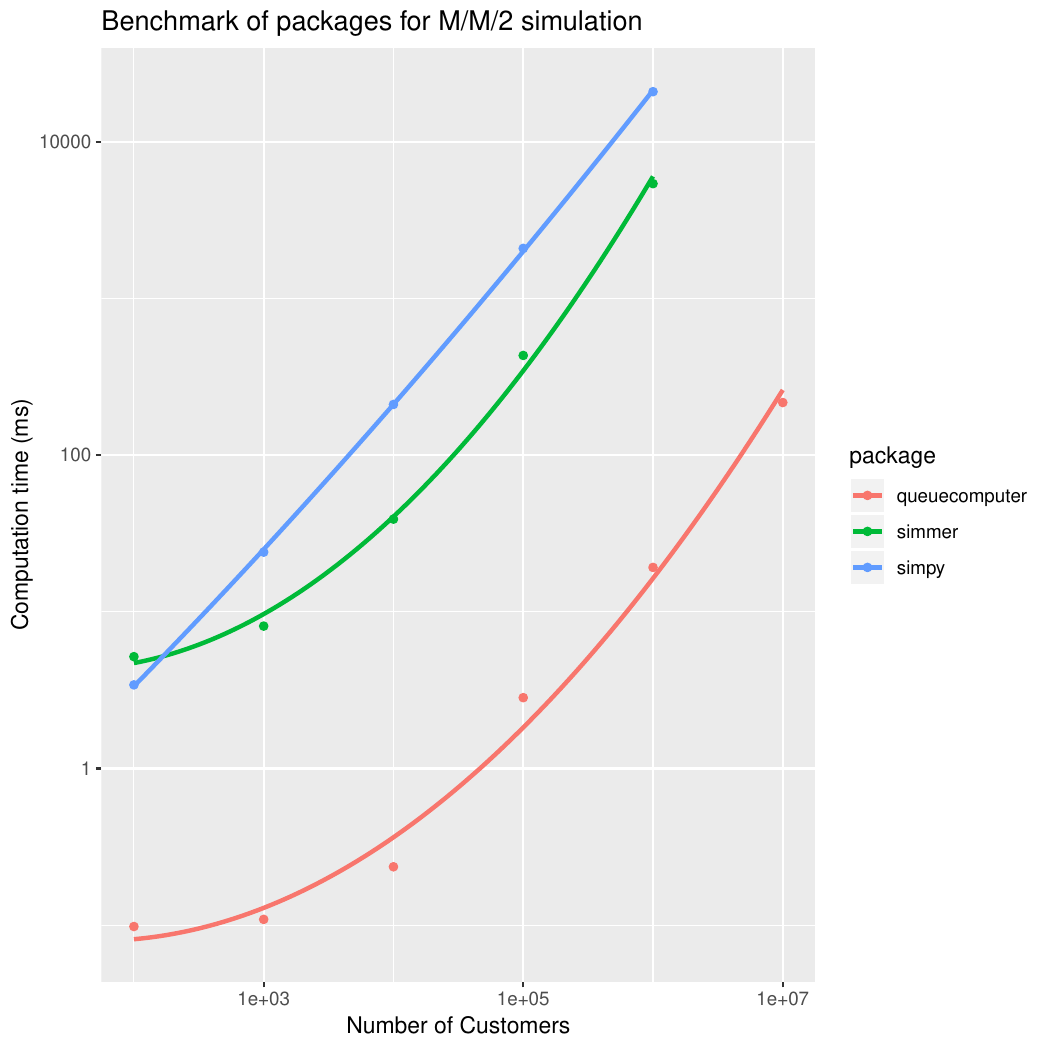}
  \caption{Computation time in milliseconds for varying numbers of passengers for each DES/queueing package. Each package returns exactly the same set of departure times since the same arrival and service times are supplied. The computation time reported here is the median time of 100 runs for each number of customers and each package. Intel (R) Core(TM) i7-6700
CPU @ 3.40GHz running Debian GNU/Linux.
 }
  \label{fig:bm_numberofpassengers}
\end{figure}

The median computation time for each package and for varying numbers of customers from $100$ to $10^6$ customers (up to $10^7$ customers for \pkg{queuecomputer}) is shown in Figure~\ref{fig:bm_numberofpassengers}. We observe phenomenal speedups for \pkg{queuecomputer} compared to both packages: compared to \pkg{simpy} speedups of 35 (at 100 customers) to 1000 (at $10^6$ customers) are observed, and for \pkg{simmer} speedups of 50 (at 100 customers) to 300 (at $10^6$ customers) are observed. The speedup is lower for smaller $n$ since \pkg{queuecomputer} approaches a minimum computation time. 

Simulating 10 million customers takes less than 1 second for \pkg{queuecomputer}. We see no reason why queues of different arrival and service distributions should not have similar speedups. This is because, as mentioned earlier, any non-negative $\mathbf{(a,s)}$ could come from two exponential distributions. 

Clearly, QDC and its implementation \pkg{queuecomputer} are a more computationally efficient way to simulate queueing systems of the form $G(t)/G(t)/K/\infty/M/FCFS$ than conventional DES algorithms implemented by \pkg{simpy} and \pkg{simmer}. 

\newpage

\section{Examples} \label{sec:Examples}

\subsection{Call center} \label{ssec:callcenter}

We demonstrate \pkg{queuecomputer} by simulating a call center. The arrival time for each customer is the time that they called, and the service time is how long it takes for their problem to be resolved once they reach an available customer service representative. Let's assume that the customers arrive by a homogeneous Poisson process over the course of the day. 

\begin{CodeChunk}
\begin{Sinput}
R> library("queuecomputer")
R> library("randomNames") 
R> library("ggplot2")
R> set.seed(1) 
R> interarrivals <- rexp(20, 1)
R> arrivals <- cumsum(interarrivals)
R> customers <- randomNames(20, name.order = "first.last")
\end{Sinput}
\end{CodeChunk}

We also need a vector of service times for every customer.

\begin{CodeChunk}
\begin{Sinput}
R> service <- rexp(20, 0.5)
R> head(service)
\end{Sinput}
\begin{Soutput}
[1] 2.6669670 1.2434810 0.4197332 0.6188957 2.2118725 1.5483755
\end{Soutput}
\end{CodeChunk}

We put the arrival and service times into the \code{queue\_step} function to compute the departure times. Here we have set the number of customer service representatives to two. The ``servers" argument is used for this input. 

\begin{CodeChunk}
\begin{Sinput}
R> queue_obj <- queue_step(arrivals, service, servers = 2, 
+    labels = customers)
R> head(queue_obj$departures_df)
\end{Sinput}
\begin{Soutput}
# A tibble: 6 x 7
  labels   arrivals service departures  waiting system_time server
  <chr>       <dbl>   <dbl>      <dbl>    <dbl>       <dbl>  <int>
1 Johatam     0.755   2.67        3.42 0.              2.67      1
2 Beatriz     1.94    1.24        3.18 0.              1.24      2
3 Devante     2.08    0.420       3.60 1.10e+ 0        1.52      2
4 Shaahira    2.22    0.619       4.04 1.20e+ 0        1.82      1
5 Ilea        2.66    2.21        5.81 9.42e- 1        3.15      2
6 Brianna     5.55    1.55        7.10 2.22e-16        1.55      1
\end{Soutput}
\end{CodeChunk}

We can see that Johatam arrives first but leaves after Beatriz. This is possible because there are two servers. Johatam's service took so long that the next two customers were served by the other server. It's easy to see how the departure times were computed in this simple example. Johatam and Beatriz were the first customers for each server so we can compute their departure time by just adding their service times to their arrival times.

\begin{CodeChunk}
\begin{Sinput}
R> firstcustomers <- arrivals[1:2] + service[1:2]
R> firstcustomers
 \end{Sinput}
 \begin{Soutput}
[1] 3.422149 3.180306
 \end{Soutput}
\end{CodeChunk}

Devante, however, had to wait for an available server, since he arrived after the first two customers arrived but before the first two customers departed. He must wait until one of these customers departs before he can be served. We add the departure time of the first customer of server 2 (Beatriz) to his service time to compute his departure time.

\begin{CodeChunk}
\begin{Sinput}
R> firstcustomers[2] + service[3]
\end{Sinput}
\begin{Soutput}
[1] 3.600039
\end{Soutput}
\end{CodeChunk}

So the first two customers had no waiting time, but Devante had to wait for an available server. We can compute the waiting times for all three customers in this manner:

\begin{CodeChunk}
\begin{Sinput}
R> depart(queue_obj)[1:3] - arrivals[1:3] - service[1:3]
\end{Sinput}
\begin{Soutput}
[1] 0.000000 0.000000 1.097774
\end{Soutput}
\end{CodeChunk}

The \code{depart} function is a convenience function for retrieving the departure times from a \code{queue\_list} object. The \code{queue\_step} function returns a \code{queue\_list} object. There is a summary method for this object within the \pkg{queuecomputer} package, this can be accessed by calling \code{summary(departures)}. 

\begin{CodeChunk}
\begin{Sinput}
R> summary(queue_obj)
\end{Sinput}
\begin{Soutput}
Total customers:
 20
Missed customers:
 0
Mean waiting time:
 1.15
Mean response time:
 3.69
Utilization factor:
 0.834333359602573
Mean queue length:
 0.858
Mean number of customers in system:
 2.42
\end{Soutput}
\end{CodeChunk}

\begin{figure}[!htb]
\centering
\includegraphics[width = 0.6\textwidth]{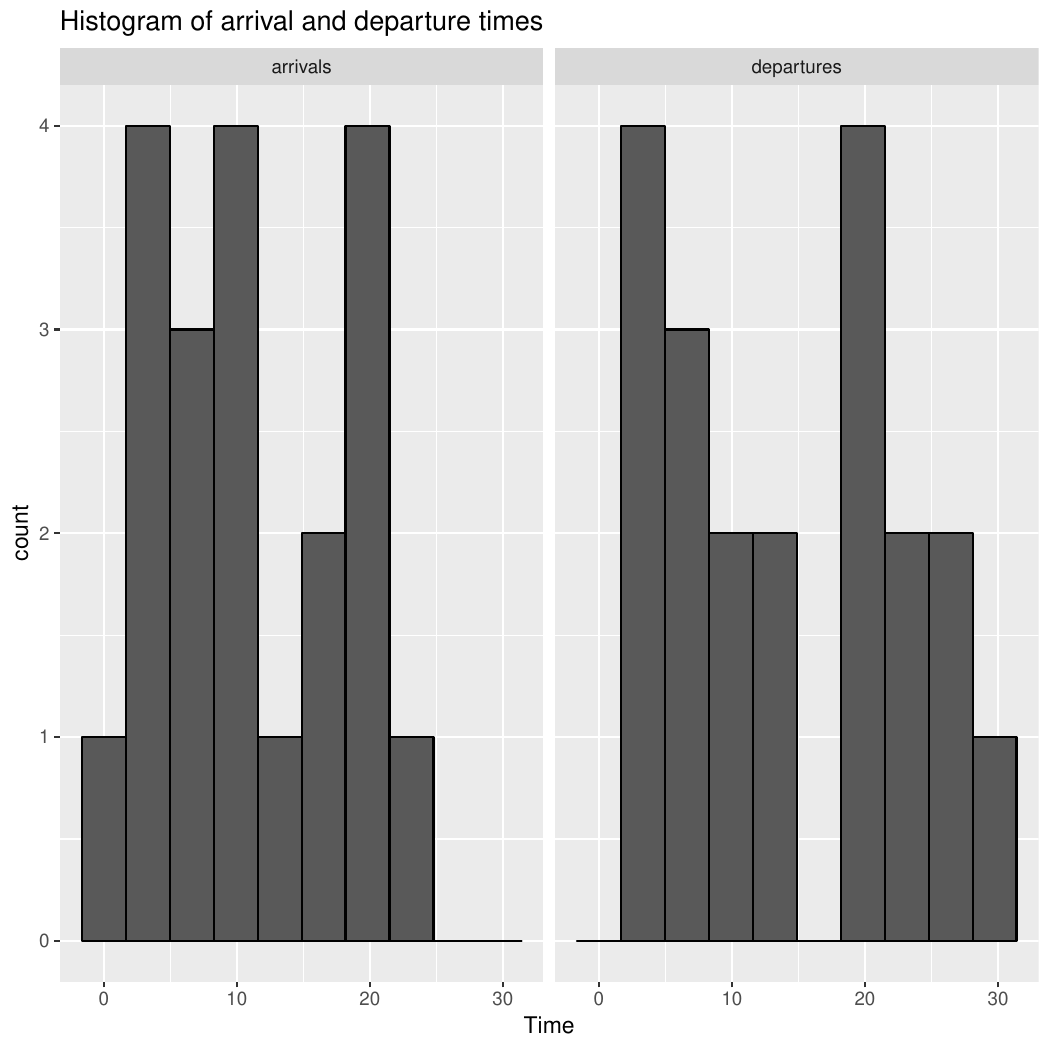}
\caption{Histogram of arrival and departure times for all customers.}
\label{fig:histogram}
\end{figure}

\begin{figure}[!htb]
\centering
\includegraphics[width = 0.6\textwidth]{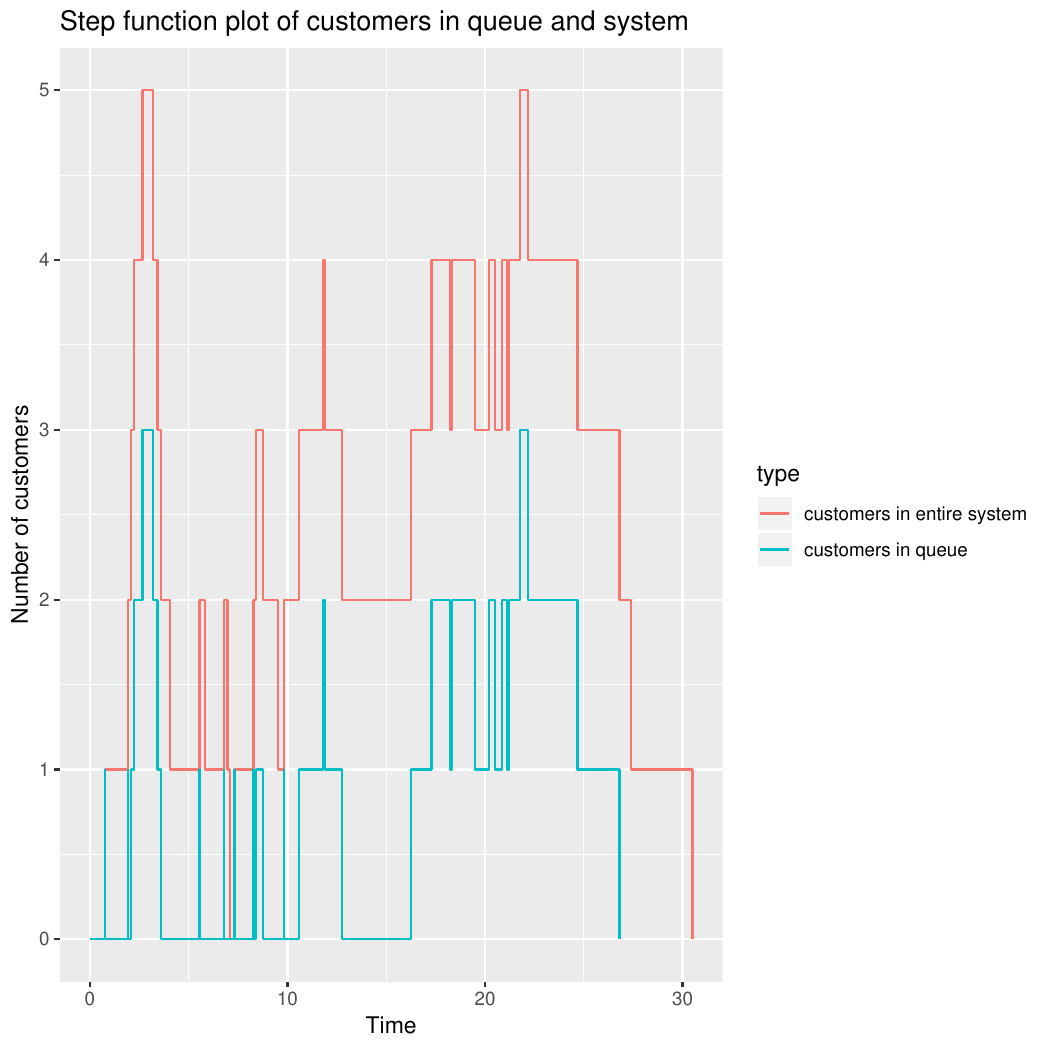}
\caption{Plot of queue length and number of customers in system over time.}
\label{fig:qlength}
\end{figure}

\begin{figure}[!htb]
\centering
\includegraphics[width = 0.6\textwidth]{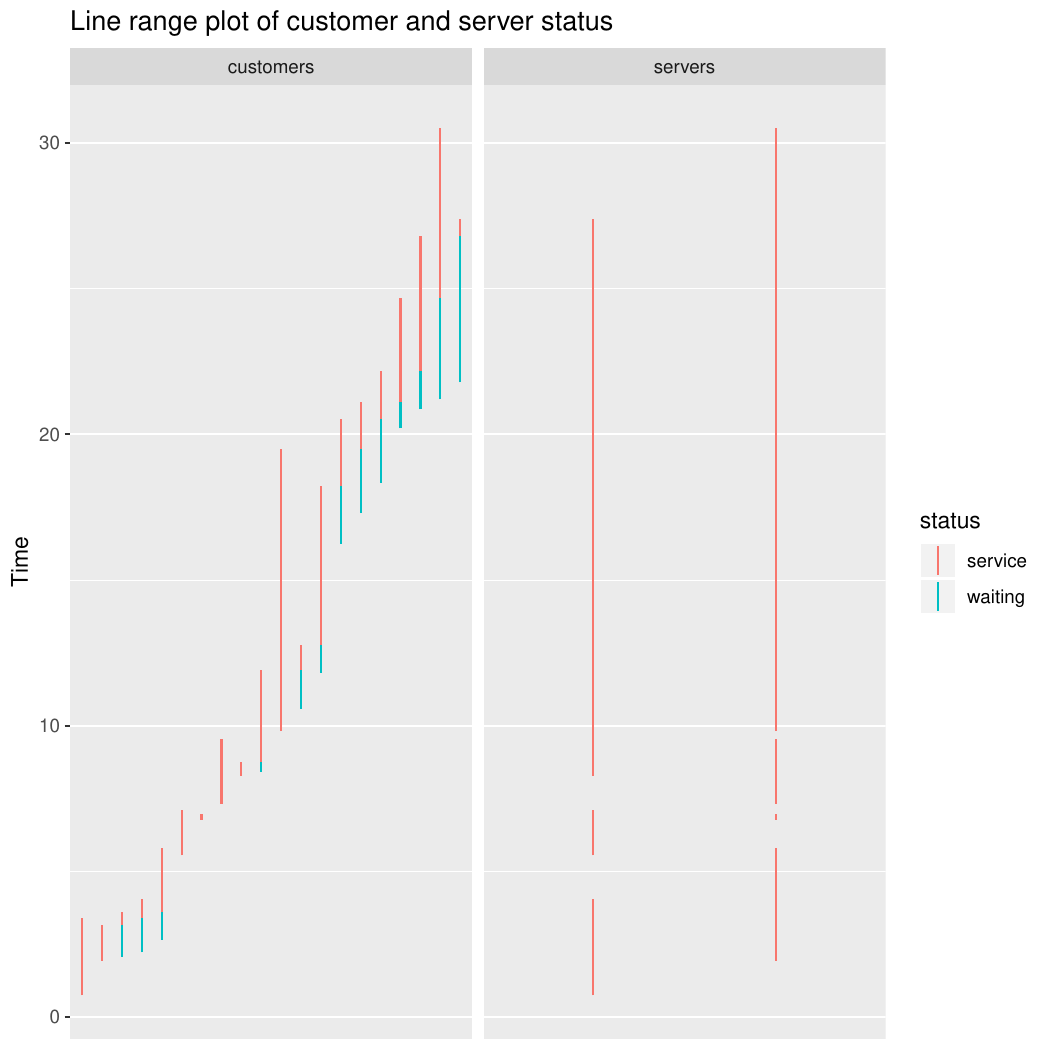}
\caption{Waiting and service times for each customer. }
\label{fig:customers}
\end{figure}

\begin{figure}[!htb]
\centering
\includegraphics[width = 0.6\textwidth]{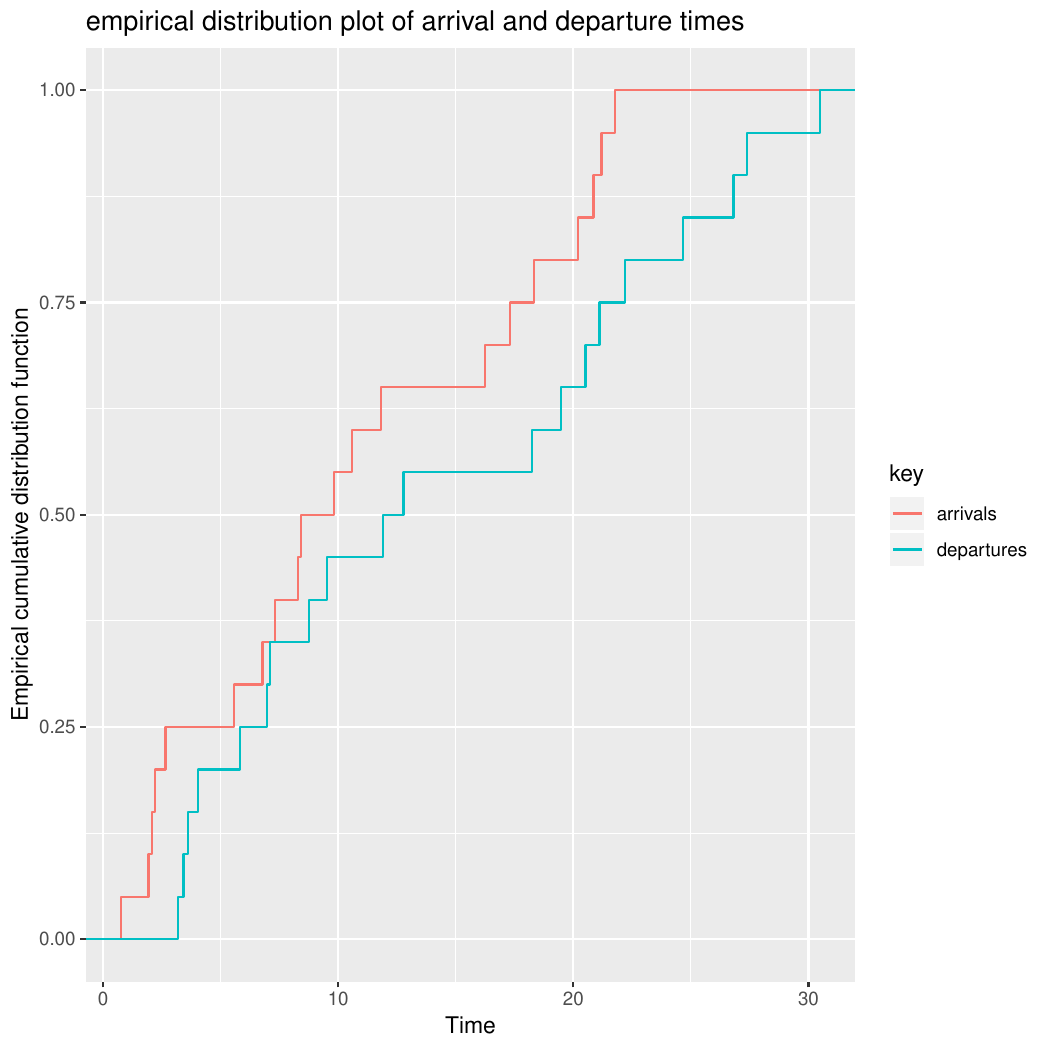}
\caption{Empirical cumulative distribution functions for arrival and departure times. For each time the different of the functions is equal to the number of customers currently in the system (in queue and currently being served).}
\label{fig:ecdf}
\end{figure}

The plot method in \pkg{queuecomputer} for \code{queue\_list} objects uses the plotting package \pkg{ggplot2} \citep{Rpkg_ggplot2} to return a list of plots. We produce four plots: a histogram of the arrival and departure times (Figure \ref{fig:histogram}); a plot of the queue length and number of customers in the system over time (Figure \ref{fig:qlength}); a plot of the waiting and service times for each customer (Figure \ref{fig:customers}); and a plot of the empirical cumulative distribution function for arrival and departure times (Figure \ref{fig:ecdf}). These plots correspond to selections $2$, $5$ and $6$ in the \code{which} argument, a similar API to the \code{plot.lm} method in the \pkg{stats} package \citep{Rproglang}.

\begin{CodeChunk}
\begin{Sinput}
R> plot(queue_obj, which = c(2, 4, 5, 6)
\end{Sinput}
\end{CodeChunk}

Notice that in Figure~\ref{fig:customers}, if we draw a horizontal line anywhere on the plot it will never pass through more than one green bar or more than one blue bar. This must be the case otherwise a server would be serving more than one customer at a time.

\subsection{International airport terminal} \label{ssec:largerairport}

\begin{figure}[!htb]
\centering
\includegraphics[width = 0.8\textwidth]{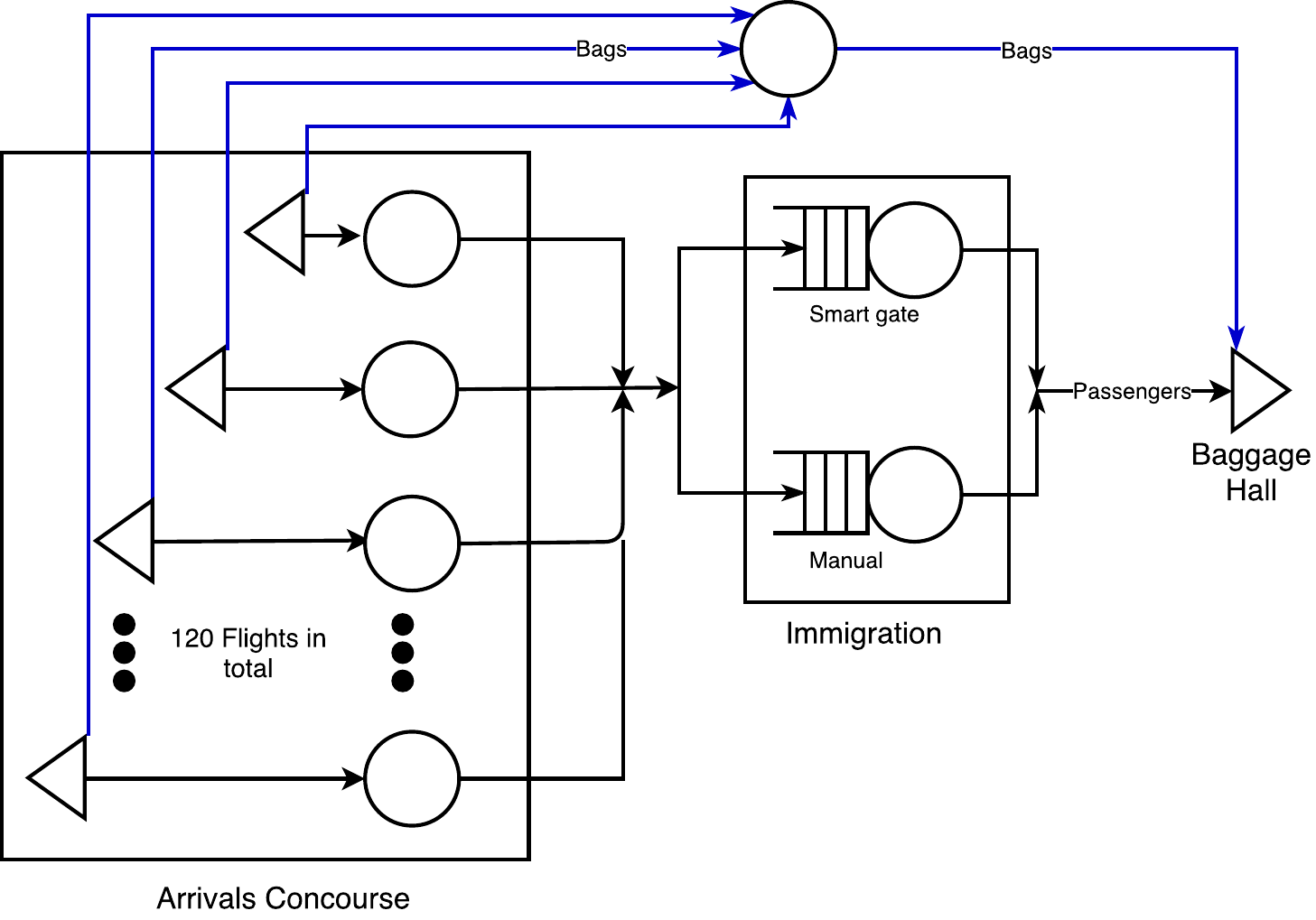}
\caption{Diagram of larger airport scenario, there are 120 flights in total and two multiserver queueing systems operate in parallel. Passengers are preassigned to travel through either the ``manual" or ``smart gate" route through immigration. The passengers and bags are ``forked'' when each aircraft arrives are then ``joined'' at the baggage hall. }
\label{fig:larger_airport}
\end{figure}

The package integrates naturally with the popular data manipulation \proglang{R} package \pkg{dplyr} \citep{Rpkg_dplyr}. We demonstrate how to integrate \pkg{queuecomputer} and \pkg{dplyr} with a more complex Airport Terminal than before (Figure~\ref{fig:larger_airport}). Passengers from a set of 120 flights disembark at the arrivals concourse and proceed through immigration using either the ``smart gate" or the ``manual gate" route, we therefore have two queues in parallel. The route taken (smart gate or manual gate) by each passenger is predetermined, but the server used by the passenger within these separate queueing systems is not. 

Their bags are unloaded from the flights and proceed to the baggage hall with a delay, the division of a passengers and bags is a fork/join network. The bags and passengers are forked at the arrival concourse and joined at the baggage hall. After immigration, the passengers proceed on to the baggage hall where they pick up their bags. 

We have a synthetic dataset of passengers \code{ID} from 120 flights \code{FlightNo}, with an average of 103.8 passengers per flight for a total of 20,758 passengers. The dataset includes (for each passenger \code{ID}): their flight number \code{FlightNo}, the arrival time of that flight \code{arrival}, the route take (smart/manual gate) by that passenger \code{route\_imm}, the arrival times to immigration after they walk through the terminal \code{arrival_imm} and the service time needed by the passenger at their immigration queueing system \code{service\_imm}. 

\begin{CodeChunk}
\begin{Sinput}
R> Passenger_df
\end{Sinput}
\begin{Soutput}
# A tibble: 25,012 x 7
   ID            FlightNo arrival route_imm arrive_imm service_imm bag_time
   <chr>         <fct>      <dbl> <fct>          <dbl>       <dbl>    <dbl>
 1 al-Akhtar, F  ABI481      565. manual          567.      0.291      574.
 2 Mcknight, De  ABI481      565. manual          567.      0.159      574.
 3 Fountain, Na  ABI481      565. manual          567.      0.225      574.
 4 Woods, Tyrel  ABI481      565. smart ga        567.      0.182      575.
 5 Peterson, Ch  ABI481      565. smart ga        566.      0.0903     575.
 6 Ruiz, Arlen   ABI481      565. smart ga        567.      0.439      575.
 7 Quick Bear,   ABI481      565. manual          568.      0.129      575.
 8 Harmon, Bren  ABI481      565. smart ga        566.      0.306      575.
 9 Caldwell, De  ABI481      565. smart ga        567.      0.320      575.
10 Hood, Colen   ABI481      565. smart ga        567.      0.339      575.
# ... with 25,002 more rows
\end{Soutput}
\end{CodeChunk}

Immigration processing is split into two routes with the \code{route\_imm} variable. The \code{"smart gate"} route has 5 servers, whereas the \code{"manual"} route has 10 servers before time 600, 12 servers between time 600 and time 780 and 8 servers from time 780 onwards. We store this information in a new \code{dataframe} called \code{server\_df}. 

\begin{CodeChunk}
\begin{Sinput}
R> server_df <- data.frame(immigration_route = c("smart gate", "manual"))
R> server_df$servers <- 
+    list(5, as.server.stepfun(x = c(600,780), y = c(10,12,8)))
\end{Sinput}
\end{CodeChunk}

To compute the departure times from the parallel servers we use the \pkg{dplyr} function \code{group_by}. The dataset is then processed as if it has been split in two. 

\begin{CodeChunk}
\begin{Sinput}
R> Passenger_df <- left_join(Passenger_df, server_df, by = "route_imm")
R> Passenger_df <- Passenger_df %>% 
+    group_by(route_imm) %>%
+    mutate(
+      departures_imm = 
+        queue(arrive_imm, service_imm, servers = servers[[1]])
+      ) %>% 
+    ungroup() %>% 
+    mutate(departures_bc = pmax.int(departures_imm, bag_time))
R> Passenger_df %>% 
+    select(FlightNo, arrive_imm, departures_imm, departures_bc)
\end{Sinput}
\begin{Soutput}
# A tibble: 25,012 x 4
   FlightNo arrive_imm departures_imm departures_bc
   <fct>         <dbl>          <dbl>         <dbl>
 1 ABI481         567.           579.          579.
 2 ABI481         567.           579.          579.
 3 ABI481         567.           580.          580.
 4 ABI481         567.           572.          575.
 5 ABI481         566.           570.          575.
 6 ABI481         567.           572.          575.
 7 ABI481         568.           580.          580.
 8 ABI481         566.           571.          575.
 9 ABI481         567.           573.          575.
10 ABI481         567.           573.          575.
# ... with 25,002 more rows
\end{Soutput}
\end{CodeChunk}

The column \code{departures\_imm} represents the times at which passengers depart immigration after having been served either through the manual counter of smart gate. The column \code{departures\_bc} represents the times that customers leave with their bags from the baggage hall. Waiting times can be summarised with the \code{summarise} function from \pkg{dplyr}, here we compute summaries of waiting times for each \code{FlightNo} and immigration route \code{route\_imm} and a summary of waiting times only by \code{route\_imm}. 

\begin{CodeChunk}
\begin{Sinput}
R> Passenger_df %>% 
+    group_by(FlightNo, route_imm) %>%
+    summarise(
+      waiting_imm = mean(departures_imm - service_imm - arrive_imm), 
+      waiting_bc = mean(departures_bc - departures_imm)
+    )
\end{Sinput}
\begin{Soutput}
# A tibble: 240 x 4
# Groups:   FlightNo [?]
   FlightNo route_imm  waiting_imm waiting_bc
   <fct>    <fct>            <dbl>      <dbl>
 1 ABI481   manual          11.3         6.29
 2 ABI481   smart gate       4.96       12.3 
 3 AEB843   manual           0.850      16.6 
 4 AEB843   smart gate       1.01       16.4 
 5 ARH364   manual          12.5         3.80
 6 ARH364   smart gate       7.36        8.17
 7 BCH445   manual           1.80       13.7 
 8 BCH445   smart gate       1.44       15.1 
 9 BJN726   manual          19.5         2.06
10 BJN726   smart gate       7.21        9.75
# ... with 230 more rows
\end{Soutput}
\begin{Sinput}
R> Passenger_df %>% 
+    group_by(route_imm) %>%
+    summarise(
+      waiting_imm = mean(departures_imm - service_imm - arrive_imm), 
+      waiting_bc = mean(departures_bc - departures_imm)
+    )
\end{Sinput}
\begin{Soutput}
# A tibble: 2 x 3
  route_imm  waiting_imm waiting_bc
  <fct>            <dbl>      <dbl>
1 manual            8.25       9.56
2 smart gate        4.49      12.8
\end{Soutput}
\end{CodeChunk}

\begin{figure}[!htb]
\centering
\includegraphics[width = 0.75\textwidth]{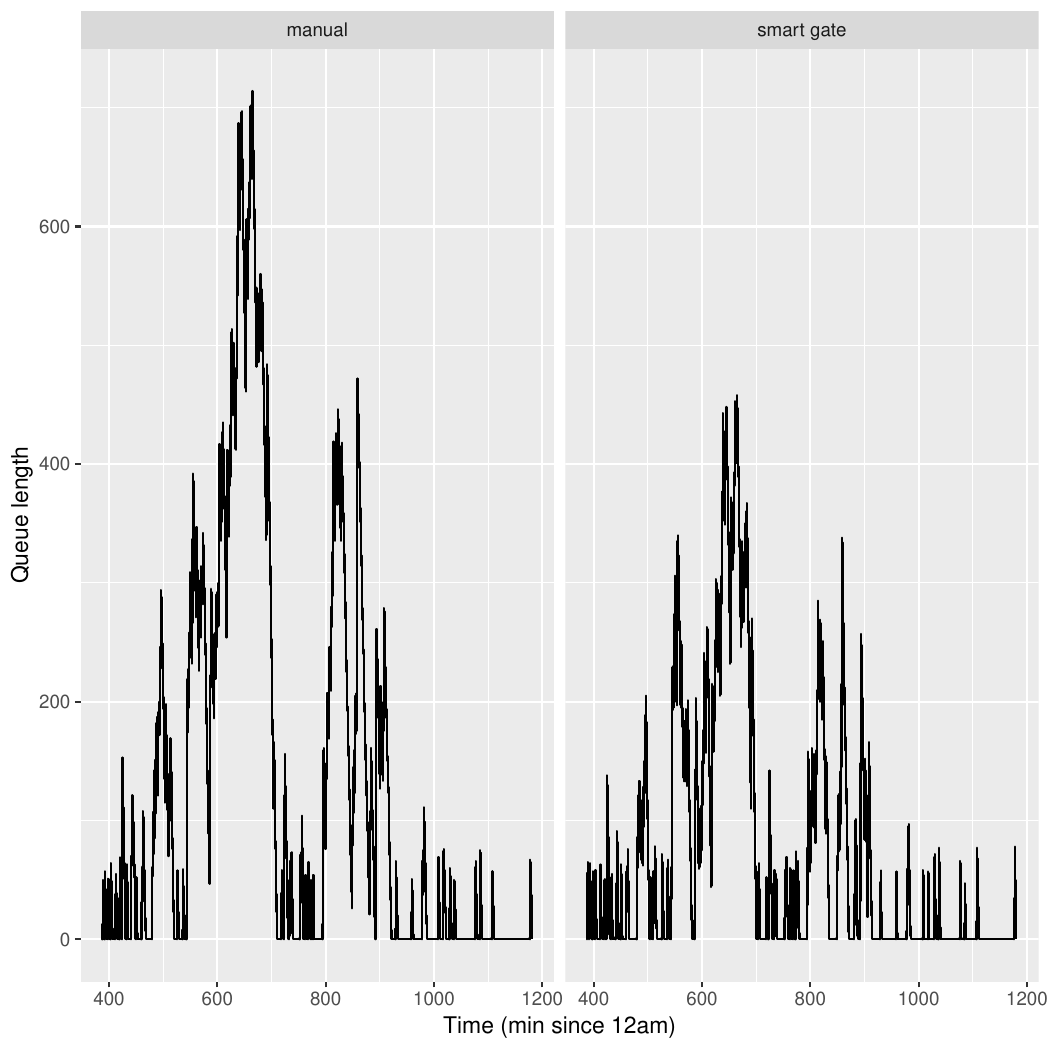}
\caption{Queue lengths over the course of the day for ``manual" and ``smart gate" immigration routes. }
\label{fig:queuelength_dplyr}
\end{figure}

\begin{figure}[!htb]
\centering
\includegraphics[width = 0.7\textwidth]{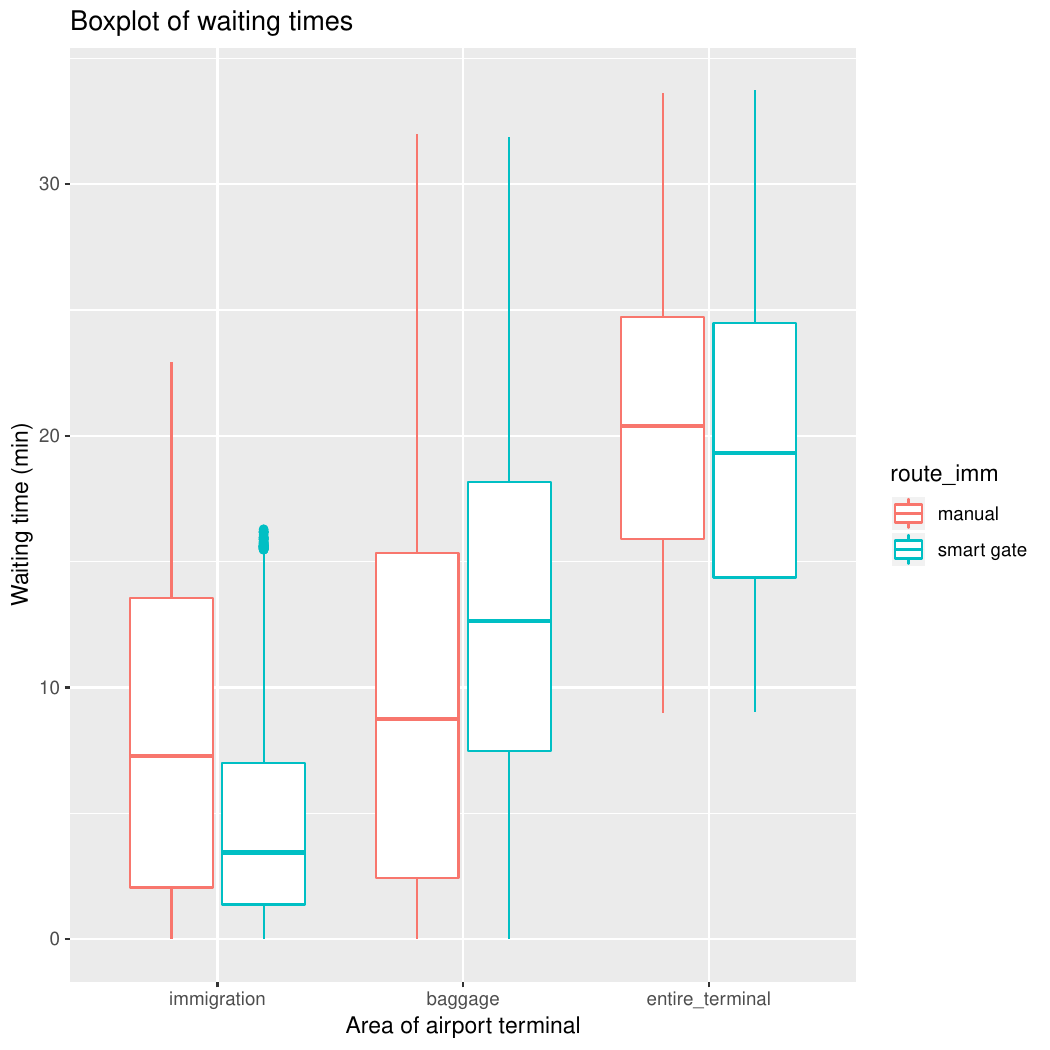}
\caption{Boxplot of waiting times for each stage of passenger processing within the international airport terminal. }
\label{fig:boxplot_dplyr.pdf}
\end{figure}

We can quickly build a complex dynamic queueing model involving tandem, parallel and fork/join topologies. The model is efficient to compute, modular and easily extended. This was achieved by combining the \pkg{queuecomputer} and \pkg{dplyr} packages. 

\section{Conclusion}

The \proglang{R} package \pkg{queuecomputer} implements QDC. It can be used to simulate any queueing systems or tandem network of queueing systems of general form $G(t)/G(t)/K(t)/\infty/n/FCFS$. Fast algorithms for multiserver queueing systems have been proposed in the past \citep{krivulin_recursive_1994,sutton_inference_2010,kin_generalized_2010}. These algorithms have generated little notice, even in the cases where their computational efficiency is demonstrated \citep{kin_generalized_2010}. QDC is conceptually simpler, more efficient memory-wise and modular. 

We validated QDC with analytic results and by replicating output generated by existing DES packages \pkg{simpy} and \pkg{simmer}. We observe speedups of up to 3 orders of magnitude. The speed of the package will allow queue simulations to be embedded within ABC algorithms, which will be addressed in future work. Unlike existing DES packages, sampling and departure time computation are clearly `decoupled' and therefore allow the user to simulate queueing systems with arrival and service time distributions of arbitrary complexity. The package integrates well with the data manipulation package \pkg{dplyr} and these two packages together allow the user to quickly and easily simulate queueing networks with parallel, tandem and fork/join topologies. 

\section*{Acknowledgements}

This work is supported by the Australian Research Council Centre of Excellence for Mathematical and Statistical Frontiers (ACEMS). This work was funded through the Australian Research Council (ARC) linkage grant ``Improving the Productivity and Efficiency of Australian Airports” (LP140100282).

\bibliography{jss3041}

\begin{table}
\resizebox{0.9\textwidth}{!}{
\begin{tabular}{|l|l|p{10cm}|}
\hline
Type & Notation & Definition \\
\hline

\multirow{10}{3cm}{Queue Specification} 

& $\lambda$ & Rate parameter of exponential inter-arrival distribution $f_\delta = \text{Exp}(\lambda)$ for $M/M/K$ queue. \\
& $\mu$ & Rate parameter of exponential service distribution $f_{\mathbf{s}} = \text{Exp}(\mu)$ for $M/M/K$ queue.  \\
& $\rho := \frac{\lambda}{K \mu}$ & Traffic intensity, defined only for $M/M/K$ queues. \\ 
& $\theta_{I}$ & Parameters of arrival and service joint distribution for QDC, $(\mathbf{a}, \mathbf{s}) \sim f_{\mathbf{a}, \mathbf{s}}(\cdot | \theta_I )$. \\

& $K$ & Number of servers. \\
& $C$ & Capacity of system.  \\
& $n$ & Total number of customers. \\
& $R$ & Service discipline of queue. \\
& FCFS & First come first serve (Type of service discipline) \\
& $L$ & Number of knot locations for server changes \\

\hline

\multirow{8}{3cm}{Input/Output} 
& $\mathbf{a} = (a_1, \cdots, a_n )$ & Arrival process, where $a_i$ is the time at which the $i$th customer arrives at the queue. \\

& $\mathbf{\delta} = (\delta_1, \cdots, \delta_n )$ & Inter-arrival process, where $\delta_i$ is $a_i - a_{i-1} \quad \forall i \in 1:n$ and $\delta_1 = a_1$. \\

& $\mathbf{s} = (s_1, \cdots, s_n )$ & Service process, where $s_i$ is the service time of the $i$th customer. \\

& $\mathbf{d} = (d_1, \cdots, d_n )$ & Departure process, where $d_i$ is the time at which the $i$th customer leaves the queue after being served. \\

& $\mathbf{p} = (p_1, \cdots, p_n )$ & Server process, where $p_i$ is the server who served the $i$th customer. \\

& $\mathbf{b} = (b_1, \cdots, b_K )$ & This vector represents the time at which each server $1:K$ will next be free. We consider this vector to be the state of the system. \\ 

& $\mathbf{x} = (x_1, \cdots, x_L )$ & Change times for number of open servers. \\ 

& $\mathbf{y} = (y_1, \cdots, y_{L+1} )$ & Number of open servers in each epoch. \\

\hline

\multirow{4}{3cm}{Performance Measures}

& $N(t)$ & Number of customers in system at time $t$. \\
& $\bar{B}$ & Time average number of busy servers, referred to as ``Resource utilization". \\
& $\bar{w}$ & Average waiting time per customer. \\

\hline

\end{tabular}
}
\caption{Notation and definitions.}
\end{table}

\end{document}